# Astrophysical *S*-factors of proton radiative capture for thermonuclear reactions

S.B. Dubovichenko, A.V. Dzhazairov-Kakhramanov

Fessenkov Astrophysical Institute "NCSRT" NSA RK, 050020, Almaty, Kazakhstan
dubovichenko@mail.ru, albert-j@yandex.ru

In this review we have considered the possibility to describe the astrophysical *S*-factors of radiative capture reactions with light atomic nuclei on the basis of the potential two-cluster model by taking into account the splitting the orbital states according to Young's schemes. Within this model, interaction of the nucleon clusters is described by local two-particle potential determined by fit to the scattering data and properties of bound states of these clusters. Many-body character of the problem is taken into account under some approximation, in terms of the allowed or forbidden by the Pauli principle states in intercluster potentials. An important feature of the approach is accounting for a dependence of interaction potential between clusters on the orbital Young scheme, which determines the permutation symmetry of the nucleon system. The astrophysical *S*-factors of the radiative capture processes in the $p^2H$, $p^7Li$ and $p^{12}C$ systems are analyzed on the basis of this approach. It is shown that the approach allows one to describe quite reasonably experimental data available at low energies, when the phase shifts of cluster-cluster scattering are extracted from the data with minimal errors. In this connection the problem of experimental error decrease is exclusively urgent for the differential cross-sections of elastic scattering of light atomic nuclei at astrophysical energies and to perform a more accurate phase shift analysis. The increase in the accuracy will allow, in future, making more definite conclusions regarding the mechanisms and conditions of thermonuclear reactions, as well as understanding better their nature in general.

## 1. Introduction

The explanation of ways of the chemical element formation in the stars is one of the significant conclusions of the modern nuclear astrophysics. The nuclear doctrine of origin of elements describes the prevalence of different elements in the Universe on the basis of characteristics of these elements taking into account physical conditions in which they can be formed. In addition, the set of considering nuclear astrophysics processes allows to interpret, for example, the star luminosity on the different stages of their evolution and to describe in general outline the process of stellar evolution itself. Hereby, the nucleosynthesis questions are closely coupled, on the one part, with the questions of structure and evolution of the stars and the Universe and on the other part with the nuclear particle interaction properties [1,2].

But there are a number of complicated and till unsolved problems, which doesn't allow to formulate the complete theory of formation and evolution of the objects in the Universe now. Let's give some examples of these up-to-date unsolved problems directly connected with nuclear astrophysics and nuclear interactions, which are followed from the existing to date nuclear physics problems [1]: *The insufficiency of experimental data of the nuclear reaction cross-sections at ultralow and astrophysical energies.*

This problem consists in the impossibility, at the modern stage of the development of experimental methods, to carry out direct measurements of the cross-sections of thermonuclear reactions in the earth's conditions for energies at which they are proceeding in the stars. We will stay at this problem more particularly, but now we will illustrate the main conceptions and representations generally using for the description of the thermonuclear reactions.

The data of cross-sections or astrophysical *S*-factors of thermonuclear reactions, including radiative capture reactions and their analysis in the frame of different theoretical models, are the main source of information about nuclear processes taken place in the Sun and stars. The considerations of similar reactions are complicated by the fact that in many cases only theoretical predictions or extrapolation results can to make up deficient experimental information about characteristics of thermonuclear processes in stellar material at ultralow energies [3].

The astrophysical *S*-factor, which determines the reaction cross-section, is the main characteristic of any thermonuclear reaction, i.e. the probability of reaction behavior at vanishing energies. It can be obtained experimentally, but it is generally possible for the majority of interacting nuclei, which are taken place in thermonuclear processes at the energy range above 100 keV÷1 MeV, but for real astrophysical calculations, for example the developing of star evolution problem, the values of astrophysical *S*-factor, are required at the energy range about 0.1-100 keV, which corresponds to the temperatures in the star core on the order of $10^6$ K÷$10^9$ K.

One of methods for obtaining the astrophysical *S*-factor at zero energy, i.e. the energy on the order of 1 keV and less, is the extrapolation of its values to lower energy range where it can be determined experimentally. It is the general way which is used, first of all, after carrying out the experimental measurements of cross-section of certain thermonuclear reaction at low energy range.

The second and evidently most preferable method consists in theoretical calculations of the *S*-factor of some thermonuclear reaction on the basis of certain nuclear models [4]. However, the analysis of all thermonuclear reactions in the frame of unified theoretical point of view is quite labor-consuming problem and later we will consider only photonuclear processes with γ-quanta, specifically the radiative capture for certain light nuclei.

The general sense of usage nuclear models and theoretical methods of calculation of thermonuclear reaction characteristics consists in the following. If the certain nuclear model correctly describes the experimental data of the astrophysical *S*-factor in that energy range where these data exist, for example 100 keV÷1 MeV, then it is reasonably assume that this model will describe the form of the *S*-factor correctly at the most low energies (about 1 keV) too.

This is the certain advantage of the approach stated above over the simple data extrapolation to zero energy, because the using model has, as a rule, the certain microscopic justification with a view to the general principles of nuclear physics and quantum mechanics.

As for the model choice, one of these models, which we use in present calculations, is the potential cluster model (PCM) of the light atomic nuclei with the classification according to the Young schemes. The model, in the certain cases, contains the forbidden states (FS) for intercluster interactions and in the simplest form gives a lot of possibilities for carrying out the similar calculations [5].

The PCM model used here is based on the assumption that the nuclei under consideration consist of two clusters. We have chosen potential cluster model because of the fact that the probability of formation of isolated nucleon associations and their isolation in the majority of light atomic nuclei is relatively high. It is confirmed by numerous experimental data and theoretical results obtained over the last fifty years [5,6]. Thus, the one-channel potential cluster model is a good approximation to the situation really existing in the atomic nucleus in many cases and for various light nuclei.

Let us emphasize the general way, which leads to the real results in the calculations of the astrophysical *S*-factor of the certain thermonuclear reaction with γ-quanta, it is the radiative capture reaction in this case. For carrying out such calculations it is necessary to have the certain data and execute following steps:

*1.     Have at one's own disposal the experimental data of the differential cross-sections or excitation functions $\sigma_{exp}$ for the elastic scattering of the considering nuclear particles (for example - $p^2H$) at lowest energies known at the present moment.*
*2.     Carry out the phase shift analysis of these data or to have the results of the phase shift analysis of the similar data that were done earlier, i.e. to know the phase shifts $\delta_L(E)$ of the elastic scattering depended on the energy E. It is one of the major parts of the whole calculation procedure of the astrophysical S-factors in PCM with FS, since it allows to obtain the potentials of the intercluster interaction.*
*3.     Construct the interaction potentials V(r) (for example for $p^2H$ system) according to the discovered phase shifts of scattering. This procedure is called as the potential description of the phase shifts of the elastic scattering in PCM with FS and it is necessary to carry out it at lowest energies.*
*4.     It is possible to carry out the total cross-sections of the photodecay process (for example $^3He + \gamma \rightarrow p + {}^2H$) and the total cross-sections of the radiative capture ($p + {}^2H \rightarrow {}^3He + \gamma$) process connected with the previous by the principle of detailed balancing, if we have the intercluster potentials obtained in such a way, i.e. to obtain the total theoretical cross-sections $\sigma(E)$ of the photonuclear reactions.*
*5.     Then, it is possible to calculate the astrophysical S-factor of the thermonuclear reaction, for example $p + {}^2H \rightarrow {}^3He + \gamma$, only if you have the total cross-sections of the radiative capture, i.e. the S(E) value as the function of energy E, at any lowest energies.*

Let's note that as of today the experimental measurements were done only for the astrophysical *S*-factor of the radiative $p^2H$ capture down to 2.5 keV, i.e. in the energy range which can be named as astrophysical. For all other nuclear systems taking part in the thermonuclear processes such measurements were thoroughly done only down to 50 keV at the best, as it was done, for example, for the $p^3H$ system.

Schematically, all these steps can be represented in the next form:

$$\sigma_{exp} \rightarrow \delta_L(E) \rightarrow V(r) \rightarrow \sigma(E) \rightarrow S(E).$$

The way, stated above, is used in this review and identical for all photonuclear reactions is independent of, for example, reaction energy or some other factors, and is the general at the consideration of any thermonuclear reaction with γ-quanta, if it is analyzed in the frame of potential cluster model with FS.

**2. Calculation methods**

*2.1. Cluster model*

The considered potential cluster model is very simple in application, since technically it is reduced to solution of the two-body problem, or, which is equivalent, to the problem of one body in the field of a force center. Therefore, an objection can be put forward that this model is absolutely inadequate to the many-body problem to which the problem of description of properties of the system consisting of $A$ nucleons is related.

In this regard it should be noted that one of the successful models in the theory of atomic nucleus is the model of nuclear shells (SM) that mathematically represents the problem of one body in the field of a force center. The physical grounds of the potential cluster model considered here should be sought in the shell model or, more precisely, in a surprising connection between the shell model and the cluster model, which is mentioned in the literature as the nucleon association model [5].

The NAM and PCM wave functions of the nucleus consisting of two clusters with the numbers of nucleons $A_1$ and $A_2$ ($A = A_1 + A_2$) have the form of antisymmetrized product of totally antisymmetric internal wave functions of clusters $\Psi(1,\ldots A_1) = \Psi(R_1)$ and $\Psi(A_1+1,\ldots,A) = \Psi(R_2)$ multiplied by the wave function of their relative motion $\Phi(R = R_1 - R_2)$,

$$\Psi = \hat{A}\ \{\Psi(R_1)\Psi(R_2)\Phi(R)\}\ ,$$

where $\hat{A}$ is the operator of antisymmetrization under permutations of nucleons belonging to different clusters. $R$ – intercluster distance, $R_1$ and $R_2$ – radius vectors of cluster center-of-mass position.

Usually cluster wave functions are chosen in such a way that they correspond to ground states of nuclei consisting of $A_1$ and $A_2$ nucleons. These wave functions are characterized by specific quantum numbers, including Young's schemes $\{f\}$, which determine the permutation symmetry of the spatial part of the wave function.

The conception of Pauli-forbidden states [5] for which total wave functions $\Psi$ with FS relative motion wave functions become zero under antisymmetrization by all $A$ nucleons is introduced in this cluster model. The potential ground i.e. really existent bound state of this cluster system is described, in the general case, by the wave function with nonzero number of nodes. We will use the Young's scheme technique for WF node number determination which we will set in below and which will be used under consideration of different cluster systems.

Thus, the idea of Pauli-forbidden states makes it possible to take into account the many-body character of the problem in terms of interaction potential between clusters [5]. In this case in practice the interaction potential is chosen in such a way that experimental data (scattering phase shifts) on elastic cluster scattering are described in the corresponding $L$ partial wave and preferably in the state with one particular Young's scheme $\{f\}$ for the spatial part of the wave function of $A$ nucleons.

Since the results of phase shift analysis in the limited energy range, as a rule, prevent unambiguous reconstruction of the interaction potential, the additional constraint on the potential is the requirement of reproduction of binding energy of the nucleus in the corresponding cluster channel. As well as description of some other static nuclear properties such as, for example, charge radius and asymptotic constant when cluster characteristics are identified with characteristics of corresponding free nuclei. This additional requirement, obviously, is an idealization, since it assumes that, in the ground state, the nucleus is 100% clusterized. Actually, the success of this potential model in description of a system of $A$ nucleons in the bound state is determined by the actual degree of clusterization of the ground state.

In this model *NN* interaction is manifested, similarly to the shell model, in creation of the mean nuclear field, and provides clusterization of the nucleus. The remaining "work" on formation of the necessary number of nodes of the wave function of relative cluster motion is executed by the Pauli principle. Therefore, it should be expected that the domain of applicability of the considered model is limited by nuclei with pronounced cluster properties.

However, some particular nuclear characteristics, even not cluster, can be mainly determined by one specific cluster channel and the small contribution of other possible cluster configurations. In this case the applied single-channel cluster model makes it possible to identify the dominating cluster channel and separate those properties of the cluster system that are determined by this channel.

*2.2. Astrophysical S-factors*

The formula for the astrophysical S-factor of the radiative capture process is of the form [7]

$$S(EJ) = \sigma(EJ) E_{cm} \exp\left(\frac{31.335 \, Z_1 Z_2 \sqrt{\mu}}{\sqrt{E_{cm}}}\right), \quad (1)$$

where $\sigma$ is the total cross-section (barn), $E_{cm}$ is the center-of-mass energy of particles, $\mu$ is the reduced mass of input channel particles (atomic mass unit) and $Z_{1,2}$ are the particle charges in elementary charge units. The numerical coefficient 31.335 that received on the basis of up-to-date values of fundamental constants, which are given in [8].

The total cross-sections of radiative capture for electric *EJ(L)* transitions in a cluster model caused by the orbital part of electric operator are given, for example, in works [9] or [10] and may be written as

$$\sigma(E) = \sum_{J, J_f} \sigma(EJ, J_f), \quad (2)$$

$$\sigma(EJ, J_f) = \frac{8\pi K e^2}{\hbar^2 q^3} \frac{\mu}{(2S_1+1)(2S_2+1)} \frac{J+1}{J[(2J+1)!!]^2} A_J^2(K) \sum_{L_i, J_i} |P_J(EJ, J_f) I_J|^2,$$

where

$$P_J^2(EJ, J_f) = \delta_{S_i S_f} (2J+1)(2L_i+1)(2J_i+1)(2J_f+1)(L_i 0 J 0 | L_f 0)^2 \begin{Bmatrix} L_i & S & J_i \\ J_f & J & L_f \end{Bmatrix}^2,$$

$$A_J(K) = K^J \mu^J \left(\frac{Z_1}{M_1^J} + (-1)^J \frac{Z_2}{M_2^J}\right),$$

$$I_J = \langle L_f J_f | R^J | L_i J_i \rangle. \quad (3)$$

Here, μ is the reduced mass and $q$ is the wave number of input channel particles; $L_f$, $L_i$, $J_f$, $J_i$ are particle momenta for input ($i$) and output ($f$) channels; $S_1$, $S_2$ - spins; $M_{1,2}$, $Z_{1,2}$, are masses and charges of input channel particles, equations (1 or 2); $K^J$, $J$ - the wave number and the momentum of γ-quanta; $I_J$ is the integral taken over wave functions of initial and final states, that is functions of the relative cluster motion from the intercluster distance $R$. Sometimes, the spectroscopic factor $S_{Jf}$ of the final state is used in the given formulas for cross-sections, but it is equal to one in the potential cluster model that we used, as it is done in work [9].

Using the formula from [11] for the magnetic transition $M1(S)$ caused by the spin part of the magnetic operator we can obtain

$$P_1^2(M1) = \delta_{S_i S_f} \delta_{L_i L_f} S(S+1)(2S+1)(2J_i+1)(2J_f+1) \begin{Bmatrix} S & L & J_i \\ J_f & 1 & S \end{Bmatrix}^2,$$

$$A_1(M1,K) = i\frac{e\hbar}{m_0 c} K\sqrt{3}\left[\mu_1 \frac{m_2}{m} - \mu_2 \frac{m_1}{m}\right], \quad (4)$$

$$I_1 = \langle \Phi_f | \Phi_i \rangle,$$

where $\mu_1$ and $\mu_2$ are magnetic momenta of proton and $^2H$, which are taken from [12] ($\mu_H$=0.857 and $\mu_p$=2.793).

The expression in square brackets in expression (4) for $A_1(M1, K)$ has been obtained on the assumption that, in the general form, for the spin part of the magnetic operator [13],

$$W_{Jm}(S) = i\frac{e\hbar}{m_0 c} K^J \sum_i \mu_i \hat{\vec{S}}_i \cdot \vec{\nabla}_i (r_i^J Y_{Jm}(\Omega_i))$$

summing over $r_i$, i.e., coordinates of centers of mass of clusters relatively to the common mass center of nucleus, is performed before the operator $\nabla$ acts on the expression in round brackets $(r_i^J Y_{Jm}(\Omega_i))$ resulting in reduction of $r_i$ degree [11],

$$\vec{\nabla}_i (r_i^J Y_{Jm}(\Omega_i)) = \sqrt{J(2J+1)} r_i^{J-1} \vec{Y}_{Jm}^{J-1}(\Omega_i).$$

In this case the coordinates $r_i$ are $R_1 = m_2/mR$ and $R_2 = -m_1/mR$, where $R$ is the relative intercluster distance and $R_1$ and $R_2$ are the distances from the common center of mass to the centers of mass of each cluster.

The electromagnetic transition operator for interaction of radiation with matter is well known in electromagnetic processes like radiative capture and photodecay. Therefore, there is a good opportunity to clarify the form of strong interaction of two particles in input channel when they are in the continuous spectra and bound states of the same particles in output channel i.e. in their discrete spectra states.

*2.3. Potentials and functions*

The intercluster interaction potentials for each partial wave, i.e., for the given orbital angular momentum *L*, and point-like Coulomb term were represented as (only nuclear part of potential is given below)

$$V(R)=V_0\exp(-\alpha R^2)+V_1\exp(-\gamma R) \qquad (5)$$

or

$$V(R)=V_0\exp(-\alpha R^2). \qquad (6)$$

Here, $V_1$ and $V_0$ are expressed in MeV, $\alpha$ and $\gamma$ have dimensions of fm$^{-2}$ and fm$^{-1}$. These are the potential parameters found from experimental data under the constraint of best description of elastic scattering phase shifts extracted in the course of phase shift analysis received from the differential cross-sections i.e. from angular distributions or excitation functions.

The expansion of WF of relative cluster motion in nonorthogonal Gaussian basis and the independent variation of parameters [10] are used by us in the two-particle variational method (VM).

$$\Phi_L(R)=\frac{\chi_L(R)}{R}=R^L\sum_i C_i \exp(-\beta_i R^2), \qquad (7)$$

where $\beta_i$ and $C_i$ are the variational expansion parameters and expansion coefficients.

The behavior of the wave function of bound states (BS) at long distances is characterized by the asymptotic constant $C_W$, having a form [14]

$$\chi_L=\sqrt{2k_0}C_W W_{-\eta L+1/2}(2k_0 R), \qquad (8)$$

where $\chi_L$ is the numerical wave function of the bound state obtained from the solution of the radial Schrödinger equation and normalized to unity; *W* is the Whittaker function of the bound state which determines the asymptotic behavior of the WF and represents the solution of the same equation without nuclear potential, i.e. long distance solution; $k_0$ is the wave number determined by the channel bound energy; $\eta$ is the Coulomb parameter; *L* is the orbital momentum of the bound state.

The root-mean-square mass radius is represented as

$$R_m^2=\frac{M_1}{M}\langle r_m^2\rangle_1+\frac{M_2}{M}\langle r_m^2\rangle_2+\frac{M_1 M_2}{M^2}I_2,$$

where $M_{1,2}$ and $\langle r_m^2\rangle_{1,2}$ are the masses and square mass radii of clusters, $M=M_1+M_2$, $I_2$ - the integral

$$I_2=\langle \chi_L(R)|R^2|\chi_L(R)\rangle$$

of the $R$ inter-cluster distance and the integration is over radial WF $\chi_L(R)$ of cluster relative motion with the orbital momentum $L$. Form of this expression is similar to (3).

The root-mean-square charge radius is represented as

$$R_z^2 = \frac{Z_1}{Z}\langle r_z^2\rangle_1 + \frac{Z_2}{Z}\langle r_z^2\rangle_2 + \frac{(Z_2 M_1^2 + Z_1 M_2^2)}{ZM^2} I_2,$$

where $Z_{1,2}$ and $\langle r_z^2\rangle_{1,2}$ are the charges and square charge radii of clusters, $Z=Z_Z+Z_2$, $I_2$ - the abovementioned integral.

The wave function $\chi_L(R)$ or $|L_i J_i\rangle$ is the solution of the radial Schrödinger equation of the form

$$\chi''_L(R) + [k^2 - V(R) - V_c(R) - L(L+1)/R^2]\chi_L(R) = 0,$$

where $V(R)$ is the inter-cluster potential represented as expressions (5) or (6) (dim. fm$^{-2}$); $V_c(R)$ is the Coulomb potential; $k$ is the wave number determined by the energy $E$ of interaction particles $k^2=2\mu E/\hbar^2$; $\mu$ is the reduced mass.

Generally, all calculations are carried out by finite-difference method (FDM), which is the modification of methods [15] and take into account Coulomb interactions. The variational method with the expansion of the wave function in nonorthogonal Gaussian basis (7) [10] is used for an additional control of calculations of bound energy and WF form.

*2.4. Cluster states classification*

The states with the minimal spin in the scattering processes of some light atomic nuclei are "mixed" according to orbital Young's schemes, for example the doublet p$^2$H state [16] is mixed according to schemes {3} and {21}. At the same time, the bound forms of these states, for example, the doublet p$^2$H channel of the $^3$He nucleus is "pure" according to scheme {3}. The method of splitting of such states according to Young's schemes is suggested in works [6,16] where in all cases the "mixed" phase shift of scattering can be represented as a half-sum of "pure" phase shifts {f$_1$} and {f$_2$}

$$\delta^{\{f_1\}+\{f_2\}} = 1/2\left(\delta^{\{f_1\}} + \delta^{\{f_2\}}\right). \tag{9}$$

In this case it is considered that {f$_1$}={21} and {f$_2$}={3}, and the doublet phase shifts, derived from the experiments, are "mixed" in accordance with these two Young's schemes. If we suppose that Instead of the "pure" quartet phase shift with the symmetry {21} one can use the "pure" doublet phase shift of p$^2$H scattering with {21} symmetry. Then it is easy to find the "pure" doublet p$^2$H phase shift with {3} symmetry [16] and use it for the construction of the interaction potential "pure" according to Young's schemes which can be used for the description of the characteristics of the bound state.

Such potential allows us to consider the bound p$^2$H state of the $^3$He nucleus. Similar ratios apply to other light nuclear systems as well, and further in each specific case we will analyze the AS and FS structure for both the scattering potentials and the interactions of the ground bound states [5].

*2.5. Phase shift analysis*

Using experimental data of differential cross-sections of elastic scattering, it is always possible to find a set of phase shifts $\delta_{S,L}^J$, which can reproduce the behavior of these cross-sections with certain accuracy. Quality of description of experimental data on the basis of a certain theoretical function (functional of several variables) can be estimated by the $\chi^2$ method which is written as [17]

$$\chi^2 = \frac{1}{N}\sum_{i=1}^{N}\left[\frac{\sigma_i^t(\theta)-\sigma_i^e(\theta)}{\Delta\sigma_i^e(\theta)}\right]^2 = \frac{1}{N}\sum_{i=1}^{N}\chi_i^2, \qquad (10)$$

where $\sigma^e$ and $\sigma^t$ are experimental and theoretical (i.e. calculated for some defined values of phase shifts $\delta_{S,L}^J$ of scattering) cross-section of elastic scattering of nuclear particles for *i*-angle of scattering, $\Delta\sigma^e$ – the error of experimental cross-sections at these angles, $N$ – the number of measurements.

The less $\chi^2$ value, the better description of experimental data on the basis of the chosen phase shift of scattering set. Expressions describing the differential cross-sections represent the expansion of some functional $d\sigma(\theta)/d\Omega$ to the numerical series and it is necessary to find such variational parameters of expansion $\delta_L$ which are the best for the description of its behavior. Since the expressions for the differential cross-sections are exact, then as $L$ approaches infinity the value of $\chi^2$ must vanish to zero. This criterion is used by us for choosing a certain set of phase shifts ensuring the minimum of $\chi^2$ which could possibly be the global minimum of a multiparameter variational problem [18].

The exact mass values of the particles were taken for all our calculations [12], and the $\hbar^2/m_0$ constant was taken to be 41.4686 MeV fm$^2$. The Coulomb parameter $\eta=\mu Z_1 Z_2 e^2/(q\hbar^2)$ was represented as $\eta=3.44476 \cdot 10^{-2} Z_1 Z_2 \mu/q$, where $q$ is the wave number determined by the energy of interacting particles in the input channel (in fm$^{-1}$), $\mu$ - the reduced mass of the particles (atomic mass unit), $Z$ - the particle charges in elementary charge units. The Coulomb potential with $R_c=0$ was represented as $V_c$(MeV)$=1.439975\, Z_1 Z_2/r$, where $r$ is the distance between the input channel particles (fm).

*2.6. Generalized matrix task for eigenvalue*

For determination of the spectrum of energy eigenvalues and eigen wave functions in the variational method upon expansion of wave function in the nonorthogonal Gaussian basis (7), the generalized matrix eigenvalue problem is solved,

$$(\mathbf{H} - E\mathbf{L})C = 0, \qquad (11)$$

where $\mathbf{H}$ is the symmetric Hamiltonian matrix; $\mathbf{L}$ is the matrix of overlapping integrals, which in the case of orthogonal basis is transformed into the identity matrix

**I**; *E* are the energy eigenvalues; and *C* are the eigenvectors of the problem.

In two-body problems for light atomic nuclei with one varied parameter $\beta_i$ in variational wave function (7), this method is rather stable, however, in the three-body nuclear system, when the variational wave function is represented in the form

$$\Phi_{l,\lambda}(r,R) = r^\lambda R^l \sum_i C_i \exp(-\delta_i r^2 - \beta_i R^2) = \sum_i C_i \Phi_i ,$$

for some values of two varied parameters $\delta_i$ и $\beta_i$, the method for finding inverse matrices using Schmidt orthogonalization sometimes results in instability and overflow upon running the computer code, which is a certain problem for solution of tasks of this type.

Therefore, an alternative method for numerical solution of the generalized matrix eigenvalue problem free from the difficulties indicated above with enhanced computer performance can be proposed. Namely, initial matrix (11) is the homogeneous system of linear equations and has nontrivial solutions only if its determinant det(**H** - *E***L**) is equal to zero. For computer numerical methods, it is not necessary to decompose the matrix *L* into triangular matrices and find the new matrix **H'** and new vectors *C'* by determining inverse matrices, as it is done using standard Schmidt orthogonalization method [19]. It is possible to decompose the nondiagonal symmetric matrix (**H** - *E***L**) into triangular matrices and seek energies resulting in zero determinant using numerical methods i.e. eigen energies, in the given energy range. In the real physical problem it is not necessary to seek all eigenvalues and eigen energies, but only one or two eigenvalues for the system energy and corresponding wave functions have to be found.

Therefore, the initial matrix (**H** - *E***L**) can be decomposed into two triangular matrices using, for example, the Khaletskii method, in such a way that the main diagonal of the upper triangular matrix **V** contains units,

$$\mathbf{A} = \mathbf{H} - E\mathbf{L} = \mathbf{NV}$$

the determinant of this matrix for det(**V**) = 1 is calculated,

$$D(E) = \det(\mathbf{A}) = \det(\mathbf{N}) \det(\mathbf{V}) = \det(\mathbf{N}) = \prod_{i=1}^{m} n_{ii}$$

and the zero of this determinant is used to find the required energy eigenvalue. Here, *m* is the dimensionality of the matrices and the determinant of the triangular matrix **N** is equal to the product of its diagonal elements.

Thus, we obtain a rather simple problem of finding the zero of a functional of one variable,

$$D(E) = 0 ,$$

the solution of this problem does not present great difficulty and can be found to any accuracy, for example, using division into halves.

As a result we eliminate the necessity of finding both inverse to **V** and **N**

matrices and carry out several matrix multiplications in order to first obtain the new matrix **H'** and then the final matrix of eigenvectors ***C***. The absence of such operations, especially finding of inverse matrices, leads to computer counting rate increasing independently of code languages. This method, which seems quite obvious in numerical implementation, made it possible to obtain good stability of the algorithm for solution of the considered problem; it does not result in overflow in the course of running the computer code, since it does not require determination of inverse to **V** and **N** matrices.

## 3. Radiative $p^2H$ capture

The first process under our consideration is the radiative capture

$$p + {}^2H \rightarrow {}^3He + \gamma,$$

which is a part of proton-proton chain and gives a considerable contribution to energy efficiency of thermonuclear reactions [1] accounting for burning of the Sun and stars of our Universe. The potential barrier for interacting nuclear particles of the p-p chain is the lowest. Thus, it is the first chain of nuclear reactions which can take place at ultralow energies and star temperatures.

For this chain, the process of the radiative $p^2H$ capture is the basic process for the transition from the primary proton fusion

$$p + p \rightarrow {}^2H + e^- + \nu_e$$

to the capture reaction of two $^3He$ nuclei [20], which is one of the final processes

$$^3He + {}^3He \rightarrow {}^4He + 2p$$

in the p-p chain.

The theoretical and experimental study of the radiative $p^2H$ capture in detail is of fundamental interest not only for nuclear astrophysics, but also for nuclear physics of ultralow energies and lightest atomic nuclei [21]. That is why the experimental researches into this reaction are in progress and at the beginning of 2000[th] years the new experimental data in the range down to 2.5 keV appeared because of the LUNA European Project.

*3.1. Photoprocesses, potentials and phase shifts of scattering*

Earlier, the total cross-sections of the photoprocesses of lightest $^3He$ and $^3H$ nuclei were considered in the frame of the potential cluster model with forbidden states in our work [22]. $E1$ transitions resulting from the orbital part of the electric operator $Q_{Jm}(L)$ [10] were taken into account in these calculations of the photodecays of $^3He$ and $^3H$ nuclei into $p^2H$ and $n^2H$ channels. The values of $E2$ cross-sections and cross-sections depending on the spin part of the electric operator turned out to be several orders less.

Further, it was assumed that $E1$ electric transitions in $N^2H$ system are possible between ground "pure" (scheme {3}) $^2S$ state of $^3H$ and $^3He$ nuclei and doublet $^2P$ scattering state mixed according to Young's schemes {3}+{21} [21]. On the basis of the approach used it was possible to obtain quite reasonable results describing of the presented at that date experimental data of $^3$H and $^3$He nuclei photodecay into the cluster channels [22].

To calculate photonuclear processes in the systems under consideration the nuclear part of the potential of inter-cluster p$^2$H and n$^2$H interactions is represented as expression (5) with a point-like Coulomb potential, $V_0$ - the Gaussian attractive part, and $V_1$ - the exponential repulsive part. The potential of each partial wave was constructed so as to correctly describe the respective partial phase shift of the elastic scattering [23]. Using this concept, the potentials of the p$^2$H interaction of the scattering processes were received. The parameters of such potentials were fully given in works [10,21,22,24], and parameters for doublet states are listed in Table 1.

**Table 1.** The potentials of the p$^2$H [10,21,22] interaction in the doublet channel.

| $^{2S+1}L$, {f} | $V_0$ (MeV) | $\alpha$ (fm$^{-2}$) | $V_1$ (MeV) | $\gamma$ (fm$^{-1}$) |
|---|---|---|---|---|
| $^2S$, {3} | -34.76170133 | 0.15 | --- | --- |
| $^2P$, {3}+{21} | -10.0 | 0.16 | +0.6 | 0.1 |
| $^2S$, {3}+{21} | -55.0 | 0.2 | --- | --- |

Then, in the doublet channel mixed according to Young's schemes {3} and {21} [16], the "pure" phases (9) with scheme {3} were separated and on their basis the "pure" $^2S$ potential of the bound state of the $^3$He nucleus in the p$^2$H channel was constructed [10,21,22,24].

The calculations of the $E$1 transition [22] show that the best results for the description of the total cross-sections of the $^3$He nucleus photodecay for the γ-quanta energy range 6-28 MeV, including the maximum value at $E_\gamma$=10-13 MeV, can be found if we use the potentials with peripheric repulsion of the $^2P$-wave of the p$^2$H scattering (see Table 1) and the "pure" according to Young's schemes $^2S$-interaction of the bound state (BS) of the Gaussian form (5) with parameters

$V_0$ = -34.75 MeV, $\alpha$ = 0.15 fm$^{-2}$, $V_1$ = 0 ,

which were obtained, primarily, on the basis of the correct description of the bound energy (with the accuracy down to few keV) and the charge radius of the $^3$He nucleus. The calculations of the total cross-sections of the radiative p$^2$H capture and astrophysical S-factors were made with these potentials at the energy range down to 10 keV [10,22]. Though, at that period of time we only knew $S$-factor experimental data in the range above 150-200 keV [25].

Recently, the new experimental data on the p$^2$H capture $S$-factor in the range down to 2.5 keV appeared in [26-28]. That is why it is interesting to know if it is possible to describe the new data on the basis of the $E$1 and $M$1 transitions in the potential cluster model with the earlier obtained $^2P$-interaction of scattering and $^2S$-potential of the p$^2$H ground state (GS) of the $^3$He nucleus.

The parameters of the "pure" doublet $^2S$-potential according to Young's scheme {3} were adjusted for a more accurate description of the experimental bound energy of $^3$He nuclei in p$^2$H channel. This potential (Table 1) has become somewhat deeper than the potential we used in our work [22] and leads to a total agreement between calculated -5.4934230 MeV and experimental -5.4934230 MeV bound energies, which are obtained by using the exact mass values of particles [12]. The difference between potentials given in work [22] and in Table 1 is primarily due to using the exact mass values of particles and more accurate description of the $^3$He nucleus bound energy in the p$^2$H channel. For these computations the absolute accuracy of searching for the bound energy in our computer program based on the finite-difference method was taken to be at the level of $10^{-8}$ MeV.

The value of the $^3$He charge radius with this potential equals 2.28 fm, which is a little higher than the experimental values listed in Table 2 [12,29,30]. The experimental radii of proton and deuteron are used for these calculations and the latter is larger than the radius of the $^3$He nucleus. Thus, if the deuteron is present in the $^3$He nucleus as a cluster, it must be compressed by about 20-30% of its size in free state for a correct description of the $^3$He radius [10].

**Table 2.** Experimental masses and charge radii of light nuclei used in these calculations [12,29,30].

| Nucleus | Radius, (fm) | Mass |
|---|---|---|
| p | 0.8768(69) | 1.00727646677 |
| $^2$H | 2.1402(28) | 2.013553212724 |
| $^3$H | 1.63(3); 1.76(4); 1.81(5)<br>The average value is 1.73 | 3.0155007134 |
| $^3$He | 1.976(15); 1.93(3); 1.877(19); 1.935(30)<br>The average value is 1.93 | 3.0149322473 |
| $^4$He | 1.671(14) | 4.001506179127 |

The asymptotic constant $C_W$ with Whittaker asymptotics (8) [31] was calculated for controlling behavior of WF of BS at long distances; its value in the range of 5-20 fm equals $C_W$=2.333(3). The error given here is found by averaging the constant in the range mentioned above. The experimental data known for this constant give the values of 1.76-1.97 [32,33], which is slightly less than the value obtained here. It is possible to give results of three-body calculations [34], where a good agreement with the experiment [35] for the ratio of asymptotic constants of $^2S$ and $^2D$ waves was obtained and the value of the constant of $^2S$ wave was found to be $C_W$=1.878.

But in work [14], which is later than [32,33], the value of 2.26(9) is given for the asymptotic constant, and this is in a good agreement with our calculations. One can see from the considerable data that there is a big difference between the experimental results of asymptotic constants received in different periods. These data are in the range from 1.76 to 2.35 with the average value of 2.06.

In the cluster model the value of $C_W$ constant depends significantly on the width of the potential well and it is always possible to find other parameters of $^2S$-potential of bound state (BS), for example:

$V_0$ = -48.04680730 MeV and $\alpha$ = 0.25 fm$^{-2}$, (12)
$V_0$ = -41.55562462 MeV and $\alpha$ = 0.2 fm$^{-2}$, (13)
$V_0$ = -31.20426327 MeV and $\alpha$ = 0.125 fm$^{-2}$, (14)

which give the same value of the bound energy of $^3$He in p$^2$H channel. The first of them at distances of 5-20 fm leads to asymptotic constant $C_W$=1.945(3) and charge radius $R_{ch}$=2.18 fm, the second variant gives $C_W$=2.095(5) and $R_{ch}$=2.22 fm, the third variant - $C_W$=2.519(3) and $R_{ch}$=2.33 fm.

It can be seen from these results that the potential (12) allows obtaining the charge radius closest to the experimental values. Further reduction of the potential width could give a more accurate description of its value, but, as it will be shown later, will not allow us to describe the S-factor of the p$^2$H capture. In this sense, the slightly wider potential (13) has the minimal acceptable width of the potential well which leads to asymptotic constant almost equal to its experimental average value 2.06 and gives a possibility to describe quite well the astrophysical S-factor in a wide energy range.

The variational method (VM) [19] was used for an additional control of the accuracy of bound energy calculations for the potential from Table 1, which allowed obtaining the bound energy of -5.4934228 MeV by using independent variation of parameters and the grid having dimension 10. The asymptotic constant $C_W$ of the variational WF at distances of 5-20 fm remains at the level of 2.34(1). The variational parameters and expansion coefficients of the radial wave function for this potential having form (7) are listed in work [3].

The potential (13) was examined within the frame of VM and the same bound energy of -5.4934228 MeV was received. The variational parameters and expansion coefficients of the radial wave function also are listed in [3]. The asymptotic constant at distances of 5-20 fm turned out to be 2.09(1) and the residual error is of the order of 10$^{-13}$ [19].

For the real bound energy in this potential it is possible to use the average value -5.4934229(1) MeV with the calculation error of finding energy by two methods equal to ±0.1 eV, because the variational energy decreases as the dimension of the basis increases and gives the upper limit of the true bound energy, but the finite-difference energy increases as the size of steps decreases and the number of steps increases [19].

*3.2. Astrophysical S-factor*

In our present calculations of the astrophysical S-factor we considered the energy range of the radiative p$^2$H capture from 1 keV to 10 MeV and found the value of 0.165 eV b for the $S(E1)$-factor at 1 keV for the potentials from Table 1. The value found is slightly lower than the known data if we consider the total S-factor without splitting it into $S_s$ and $S_p$ parts resulting from M1 and E1 transitions. This splitting was made in work [27], where $S_s(0)$=0.109(10) eV b and $S_p(0)$=0.073(7) eV b. At the same time, the authors give the following values $S_0$=0.166(5) eV b and $S_1$=0.0071(4) eV b keV$^{-1}$ in the linear interpolation formula

$S(E_{c.m.}) = S_0 + E_{c.m} S_1$, (15)

and for $S(0)$ leads to the value of 0.166(14) keV b, which was received taking into account all possible errors. The results with the splitting of the $S$-factor into $M1$ and $E1$ parts are given in one of the first of works [25], where $S_s$=0.12(3) eV b, $S_p$=0.127(13) eV b for the total $S$-factor 0.25(4) eV b. The abbreviation c.m. – center-of-mass system.

As it can be seen, there is a visible difference between these results, so, in future we will generally orient to the total value of the $S$-factor at zero energy which was measured in various works. Furthermore, the new experimental data [28] lead to the value of total $S(0)$=0.216(10) eV b and this means that contributions of $M1$ and $E1$ will change as compared with [27]. The following parameters of linear extrapolation (15) are given in work [28] $S_0$=0.216(6) eV b and $S_1$=0.0059(4) eV b keV$^{-1}$, that are noticeably differ from the data of work [27].

The rest known extractions of the $S$-factor from the experimental data, without splitting to $M1$ and $E1$ parts, at zero energy give the value of 0.165(14) eV b [36]. The previous measurements by the same authors lead to the value 0.121(12) eV b [37], and for theoretical calculations of work [38] the values $S_s$=0.105 eV b, $S_p$=0.0800-0.0865 eV b are received for different models.

One can see that the cited experimental data over the last 10-15 years are very ambiguous. These results allow only to come to a conclusion that the value of total $S$-factor at zero energy is in the range 0.11-0.23 eV b. The average of these experimental measurements equals 0.17(6) eV b what is in a properly agreement with the value calculated here on the basis of the $E1$ transition only.

The dashed line in Fig. 1 shows the calculation results for $E1$ transition for the potential of the ground state (13). The total $S$-factor is shown in Fig. 1 with a solid line which demonstrates clearly the small contribution of $M1$ transition to the $S_s$-factor at energies above 100 keV and its considerable influence in the energy range of 1-10 keV.

The total $S$-factor dependence on energy in the range of 2.5-50 keV is in complete accordance with the findings of works [27,28] and for the $S_s$-factor of the $M1$ transition at 1 keV we obtained the value of 0.077 eV b, which leads to the value of 0.212(5) eV b for the total $S$-factor and which is in a good agreement with the new measurements data from LUNA project [28]. And as it can be seen from Fig. 1, at the energies of 1-3 keV the value of the total $S$-factor is more stable than it was for the $E1$ transition and we consider it to be absolutely reasonable to write the result as 0.212 eV b with the error of 0.005.

However, it is necessary to note that we are unable to build the scattering $^2S$-potential uniquely because of the ambiguities in the results of different phase shift analyses. The other variant of potential with parameters $V_0$=-35.0 MeV and α=0.1 fm$^{-2}$ [10,22], which also describes well the $S$ phase shift of elastic scattering, leads at these energies to cross-sections of the $M1$ process several times lower.

Such a big ambiguity in parameters of the $^2S$-potential of scattering, associated with errors of phase shifts extracted from the experimental data, does not allow us making certain conclusions about the contribution of the $M1$ process in the p$^2$H radiative capture. If the BS potentials are defined by the bound energy, asymptotic constant and charge radius quite uniquely and the potential description of the scattering phase shifts, which are "pure" in accordance with Young's schemes, is an additional criteria for determination of such parameters, then, for the construction of the scattering potential it is necessary to carry out a more accurate phase shift analysis for the $^2S$-wave and to take into account the spin-orbital splitting of $^2P$ phase shifts at low energies, as it was done for the elastic p$^{12}$C scattering at energies 0.2-1.2 MeV [39]. This will allow us to adjust the potential parameters used in the calculations of the p$^2$H capture in the potential cluster model.

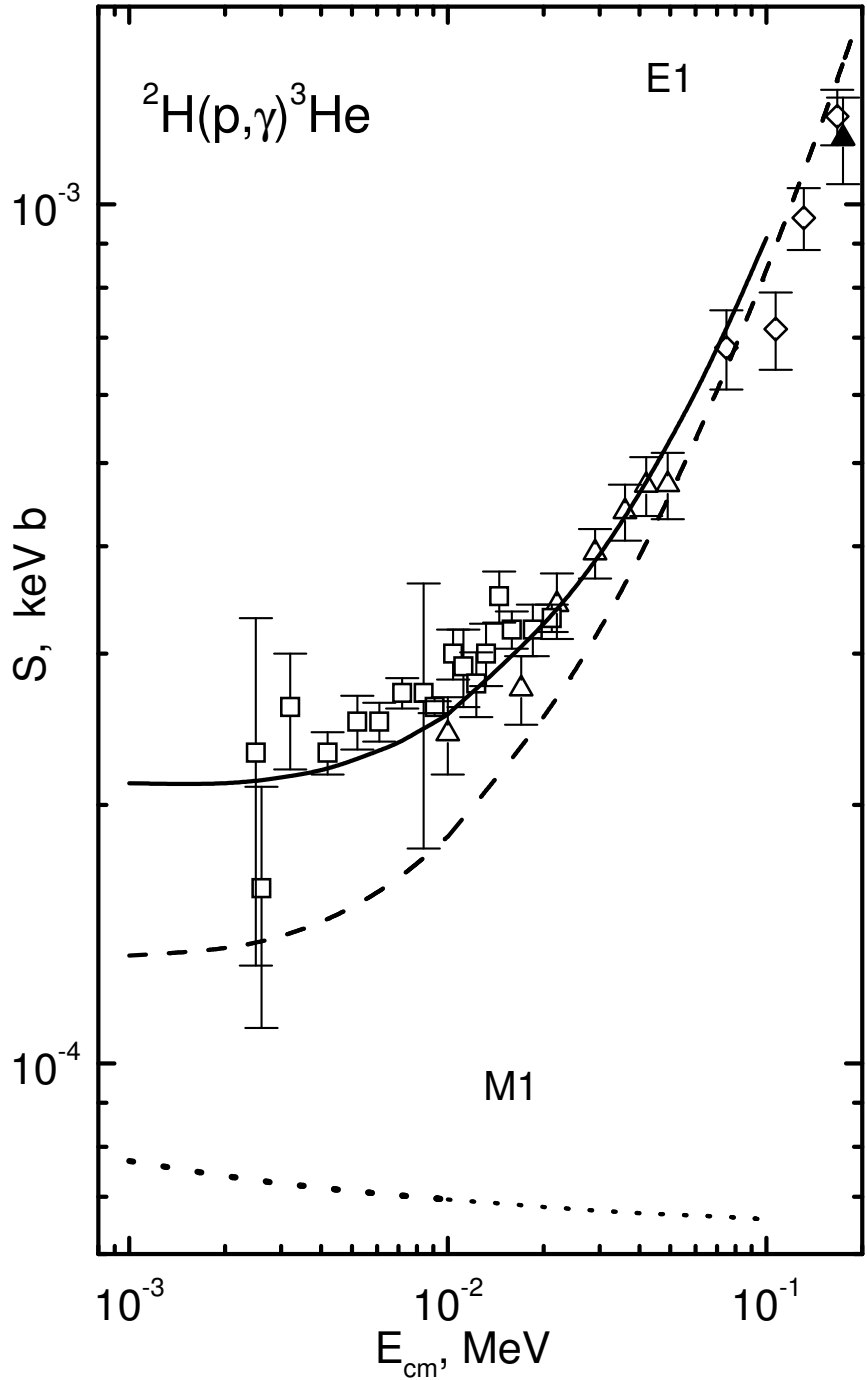

**Fig. 1.** Astrophysical *S*-factor of p$^2$H radiative capture in the range 1 keV-0.3 MeV. Lines: calculations with the potentials mentioned in the text. Triangles denote the experimental data from [25], open rhombs from [26], open triangles from [27], open squares from [28].

Thus, the *S*-factor calculations of the p$^2$H radiative capture for the *E*1 transition at the energy range down to 10 keV, which we carried out about 15 years ago [22], when the experimental data above 150-200 keV were only known, are in a good agreement with the new data of works [26-28] in the energy range 10-150 keV. And this is true about both the potential from Table 1 and the interaction with parameters from (13). The results for the two considered potentials at the energies lower than 10 keV practically fall within the error band of work [28] and show that the *S*-factor tends to remain constant at energies 1-3 keV. In our calculations [22] there was intrinsically predicted the behavior of the pd → $^3$Heγ *S*-factor in the energy range from 10-20 to 150-200 keV, which value is generally defined by the *E*1 transition at these energies.

In spite of the uncertainty of the *M*1 contribution to the process, which results from the errors and ambiguity of $^2S$-phases of scattering, the scattering potential (set forth in Table 1) with mixed Young's schemes in the $^2S$-wave allows obtaining a reasonable value for the astrophysical $S_s$-factor of the magnetic transition in the range of low energies. At the same time, the value of the total *S*-factor is in a good agreement with all known experimental measurements at energies from 2.5 keV to 10 MeV.

As a result, the PCM based on the intercluster potentials adjusted for the elastic scattering phase shifts and GS characteristics, for which the FS structure is determined by the classification of BS according to Young's orbital schemes and potential parameters suggested as early as 15 years ago [22], allows describing correctly the astrophysical *S*-factor for the whole range of energies under consideration.

## 4. Radiative p$^7$Li capture

The reaction of radiative capture

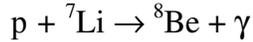

p + $^7$Li → $^8$Be + γ

at ultralow energies resulting in formation of unstable $^8$Be nucleus which decays into two α-particles may take place along with the weak process as one of the final reactions of the proton-proton chain [4]. Therefore the in-depth study of this reaction, in particular of the form and energy dependence of the astrophysical *S*-factor, is of a certain interest for the nuclear astrophysics.

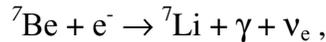

$^7$Be + e$^-$ → $^7$Li + γ + ν$_e$ ,

To calculate the astrophysical *S*-factor of radiative p$^7$Li capture in the potential cluster model [5,10] which we usually use for such calculations [40] it is necessary to know partial potentials of p$^7$Li interaction in the continuous and discrete spectra. We will again assume that such potentials should follow the classification of cluster states by orbital symmetries [5] as it was assumed in our earlier works [21,41].

As before, the potentials of scattering processes will constructed on the basis of description of elastic scattering phase shifts obtained from experimental data while the interactions in bound states are determined by the requirement to reproduce the main characteristics of the bound state of the nucleus assuming that it is mainly due to cluster channel consisting of the input particles of the reaction under consideration.

For example, in the radiative capture process the $^2$H$^4$He particles colliding at low energies form $^6$Li nucleus in the ground state and the remaining energy is released as a γ quantum. Since there is no restructuring in such reactions we can consider potentials of one and the same nuclear system of particles that is the $^2$H$^4$He system in continuous and discrete spectra. In the latter case it is assumed that the ground state of the $^6$Li nucleus is very likely caused by the cluster $^2$H$^4$He configuration. Such approach leads to quite reasonable results of the description of the astrophysical S-factors of this and some other reactions of radiative capture [40].

It seems that in this case the $^8$Be nucleus does not consist of cluster p$^7$Li system and most probably is determined by the $^4$He$^4$He configuration into which it decays. However, it is possible that the $^8$Be nucleus is in the bound state of the p$^7$Li channel for a while just after of the reaction of the radiative p$^7$Li capture and only after this it changes to the state defined by the unbound $^4$He$^4$He system. Such an assumption makes it possible to consider the $^8$Be nucleus as the cluster p$^7$Li system and use PCM methods, at least at the initial stage of its formation in the reaction p+$^7$Li → $^8$Be+γ [41].

*4.1. Classification of the orbital states*

First, we would like to note that the p$^7$Li system has the $T_z = 0$ isospin projection and it is possible for two values of total isospin $T = 1$ and 0 [42], therefore p$^7$Li channel is mixed by isospin as p$^3$H system [43], even though as it will be shown later both of isospin states ($T = 1,0$), in contrast to p$^3$H system, in the triplet spin state correspond to the allowed Young's scheme {431} [10]. The cluster channels p$^7$Be and n$^7$Li with $T_z = ±1$ and $T = 1$ are pure by isospin in a complete analogy with the p$^3$He and n$^3$H systems [44].

The spin-isospin Young's schemes of the $^8$Be nucleus for the p$^7$Li channel are the product of spin and isospin parts of the WF. Particularly, under consideration of any of these momenta we will have {44} scheme at the ground state of the $^8$Be nucleus with the momentum equals zero, scheme {53} for a certain state with momentum equals one and for the state with momentum equals two - {62} symmetry form.

If the scheme {7} is used for the $^7$Li nucleus then possible Young's schemes of p$^7$Li system turn out to be forbidden, because of the rule that there can not be more than four cells in a row [44,45], and they correspond to forbidden states with configurations {8} and {71} and relative motion momenta $L = 0$ and 1, which is determined by Elliot rule [45]. The p$^7$Li system, in the triplet spin state, contains forbidden states with the scheme {53} in $P_1$-wave and {44} in $S_1$-wave and allowed state with the configuration {431} at $L = 1$ when the scheme {431} is accepted for the $^7$Li nucleus [46].

Thus, the p$^7$Li potentials in the different partial waves should have the forbidden bound state {44} in the $S_1$-wave and forbidden and allowed bound levels in the $P_1$-wave with schemes {53} and {431}, respectively. The considered classification is true for any isospin state of the p$^7$Li system ($T = 0$ or 1) in triplet spin channel. Allowed symmetries are absent for spin $S = 2$ and all Young's schemes listed above correspond to forbidden states.

Probably, as it was in a previous case for the p$^6$Li system, it is more correctly to consider both allowed schemes {7} and {43} for bound states of the $^7$Li because of the fact that they are present in FS and AS in the $^3$H$^4$He configuration of this nucleus [40]. Then the level classification will be slightly different, the number of forbidden states will increase and an extra forbidden state will appear in each partial wave. Such more complete scheme of FS and AS states, per se, is a sum of the first and the second cases considered above is listed in Table 3.

**Table 3.** The classification of the orbital states in $p^7Li$ ($n^7Be$) systems of isospin $T=0,1$ [46]. Note: $T$, $S$ and $L$ are, respectively, the isospin, spin and orbital momentum of particles; $\{f\}_S$, $\{f\}_T$, $\{f\}_{ST}$ and $\{f\}_L$ are, respectively, the spin, isospin, spin-isospin and possible orbital Young's schemes; $\{f\}_{AS}$ and $\{f\}_{FS}$ are the Young's schemes of, respectively, allowed and forbidden states. The conjugate schemes are printed in boldface italic.

| System | T | S | $\{f\}_T$ | $\{f\}_S$ | $\{f\}_{ST} = \{f\}_S \otimes \{f\}_T$ | $\{f\}_L$ | L | $\{f\}_{AS}$ | $\{f\}_{FS}$ |
|---|---|---|---|---|---|---|---|---|---|
| $p^7Li$ $n^7Be$ | 0 | 1 | {44} | {53} | {71} + {611} + {53} + {521} + {431} + {4211} + {332} + ***{3221}*** | {8} {71} {53} {44} ***{431}*** | 0 1 1,3 0,2,4 1,2,3 | - - - - {431} | {8} {71} {53} {44} - |
| | | 2 | {44} | {62} | {62} + {521} + {44} + {431} + {422} + {3311} | {8} {71} {53} {44} {431} | 0 1 1,3 0,2,4 1,2,3 | - - - - - | {8} {71} {53} {44} {431} |
| $p^7Be$ $n^7Li$ $p^7Li$ $n^7Be$ | 1 | 1 | {53} | {53} | {8} + 2{62} + {71} + {611} + {53} + {44} + 2{521} + {5111} + {44} + {332} + 2{431} + 2{422} + {4211} + {3311} + ***{3221}*** | {8} {71} {53} {44} ***{431}*** | 0 1 1,3 0,2,4 1,2,3 | - - - - {431} | {8} {71} {53} {44} - |
| | | 2 | {53} | {62} | {71} + {62} + {611} + 2{53} + 2{521} + 2{431} + {422} + {4211} + {332} | {8} {71} {53} {44} {431} | 0 1 1,3 0,2,4 1,2,3 | - - - - - | {8} {71} {53} {44} {431} |

*4.2. Potential description of scattering phase shifts*

    Because of the isospin mixing the phase shifts of the $p^7Li$ elastic scattering are represented as a half sum of the isospin pure phase shifts [41] in complete analogy with the $p^3H$ system considered above [43,44]. The phase shifts with $T = 1,0$ mixed by isospin are usually determined as a result of the phase shift analysis of the experimental data of differential cross-sections of the elastic scattering or excitation function. The pure phase shifts with isospin $T = 1$ are determined from the phase shift analysis of the $p^7Be$ or $n^7Li$ elastic scattering. As a result it is possible to find pure $p^7Li$ phase shifts of scattering with $T = 0$ and construct the interaction model using these results which have to correspond to the potential of the bound state of the $p^7Li$ system in the $^8Be$ nucleus [47]. Just the same method of phase shift separation was used for the $p^3H$ system [44] and its absolute validity was shown [10].
    However, we failed to find experimental data of differential cross-sections or

phase shifts of the p$^7$Be or n$^7$Li elastic scattering at astrophysical energies, so here we will consider only isospin-mixed potentials of the elastic scattering processes in the p$^7$Li system and pure potentials of the bound state with $T = 0$ which are constructed on the base of description BS characteristics and are chosen in the Gaussian form with point-like Coulomb term (6).

The phase shifts of the p$^7$Li elastic scattering received from the phase shift analysis of the experimental data of excitation functions [48] taking into account spin-orbital splitting at the energies up to 2.5 MeV are given in the work [49]. These phases, which are equal to zero in the in $S_1$-wave at energies down from 600-700 keV, we will use later for the intercluster potential construction for the p$^7$Li elastic scattering in $S_1$- and $P_1$-waves. Since later we will consider the low and astrophysical energy range only, then we will limit the energy range from 0 keV to 700 keV. Practically zero phase shift at these energies is received with the potential of the form (6) and parameters:

$V_0 = -147.0$ MeV and $\alpha = 0.15$ fm$^{-2}$.

Such potential contains two FS as it follows from the state classification given above. Of course, $S_1$-phase shift at about zero one can obtain from the other variants of potential parameters with two FS. In this regard it is not possible to fix its parameters unambiguously and the other combinations of $V_0$ and $\alpha$ are possible. However, this potential, as the potential given above, should have comparatively large width which gives small phase shift change when the energy changes in the range from 0 to 700 keV.

There is an over-threshold level in the $P_1$-wave with the energy 17.640 MeV and $J^PT = 1^+1$ or 0.441 MeV in laboratory system (l.s.) which is above the threshold of the cluster p$^7$Li channel in the $^8$Be nucleus, with the bound energy of this channel being -17.2551 MeV [50]. The 0.441 MeV level has very small width of only 12.2(5) keV [50] for the p$^7$Li → $^8$Be$\gamma$ radiative capture reaction and p$^7$Li elastic scattering. Such a narrow level leads to the sharp rise of the $P_1$-phase shift of elastic scattering which should be mixed by spin states $^5P_1$ and $^3P_1$ [50] for the total moment $J = 1$. The phase shift, which is shown with points in Fig. 2 [49], is described by the Gaussian potential (6) with parameters:

$V_0 = -5862.43$ MeV and $\alpha = 3.5$ fm$^{-2}$.

This potential, mixed in isospin $T = 0$ and 1, has two FS and the calculation results of the $P_1$-phase shift of the elastic scattering are shown in Fig. 2 with a solid line. The potential parameters which describe the $P_1$-phase shift are fixed quite unambiguously at the interval of sharp increase obtained from the experimental data and the potential itself should have very small width.

Since, later we will consider astrophysical $S$-factor only at the energies from 0 to 700 keV, it can be deemed that both of the potentials received above give a good description of the results of the phase shift analysis for two considered partial waves in this energy range.

The following parameters of the potential of the bound $^3P_0$-state of the p$^7$Li system corresponding to the ground state of the $^8$Be nucleus in the examined cluster channel are obtained:

$V_0 = -433.937674$ MeV and $\alpha = 0.2$ fm$^{-2}$.

The bound energy -17.255100 MeV with the accuracy of $10^{-6}$ MeV, the root-mean-square radius is equal to 2.5 fm and the asymptotic constant, calculated with the help of Whittaker functions (8), equals $C_W = 12.4(1)$ were obtained with such a potential. The error of the constant is estimated by its averaging in the range 6-10 fm where the asymptotic constant is practically stable. In addition to the allowed BS corresponding to the ground state of the $^8$Be nucleus such $P$-potential has two FS in total correspondence with the classification of orbital cluster states.

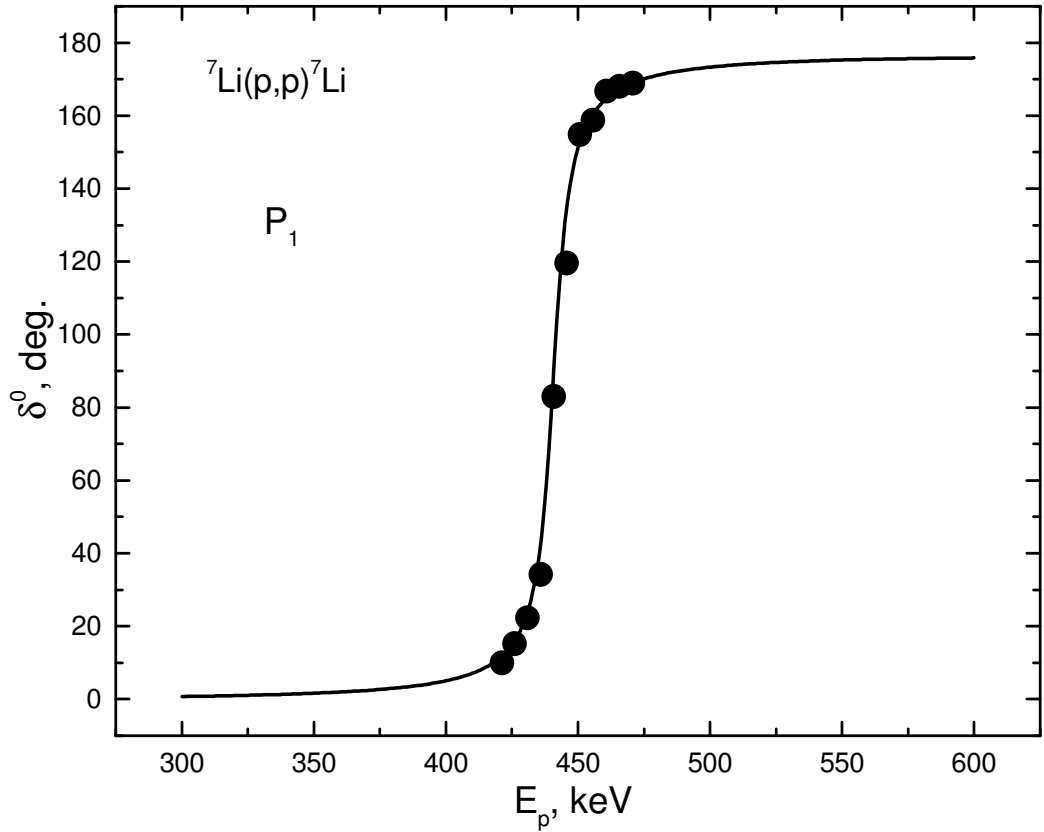

**Fig. 2.** $^5P_1$-phase mixed with $^3P_1$-phase of the elastic p$^7$Li scattering at low energies. Points - phase shifts received from the experimental data in work [49]. Line - calculations with the Gaussian potential based on parameters given in the text.

It seems that the root-mean-square radius of the $^8$Be nucleus in the cluster p$^7$Li channel should not differ a lot from the $^7$Li radius which equals 2.35(10) fm [50], since the nucleus is in a strongly bound (~ -17 MeV) i.e. compact state. Moreover, at such bound energy the $^7$Li nucleus itself can be in deformed, compressed form as it is for deuteron in the $^3$He nucleus [47]. Therefore, the value of the root-mean-square radius for the p$^7$Li channel in the GS of the $^8$Be nucleus received above has quite a reasonable value.

The variational method with the expansion of the cluster wave function of the p$^7$Li system in nonorthogonal Gaussian basis [19] is used for an additional control of the accuracy of bound energy calculations and the energy -17.255098 MeV with $N=10$ order of matrix were obtained for this potential which differ from the given above finite-difference value by 2 eV only. Residuals [19] are of the order of $10^{-11}$, asymptotic constant at the range 5-10 fm equals 12.3(2), the charge radius does not differ from previous results. Expansion parameters of the received variational GS radial wave function of the $^8$Be nucleus in the p$^7$Li cluster channel are listed in [3].

As it was told before, the variational energy decreases as the dimension of the basis increases and gives the upper limit of the true bound energy, but the finite-difference energy increases as the size of steps decreases and the number of steps increases [19], therefore it is possible to use the average value -17.255099(1) MeV for the real bound energy in this potential. Thus, the calculation error of the bound energy of the $^8$Be nucleus in the cluster p$^7$Li channel using two different methods is about ±1 eV.

*4.3. Astrophysical S-factor*

While considering electromagnetic transitions for the *S*-factor calculations we will take into account the *E*1 process from the $^3S_1$-wave of scattering to the ground bound state of the $^8$Be nucleus in the cluster p$^7$Li channel with $J^PT = 0^+0$ and the *M*1 transition from the $P_1$-wave of scattering (see Fig. 2) also to the GS of the nucleus. Cross-sections of the *E*1 transition from the $^3D_1$-wave of scattering (with potential for the $^3S_1$-wave at *L*=2) to the GS of the $^8$Be nucleus are by 2-4 orders lower than from the $^3S_1$-wave transition at the energy range 0-700 keV. Further on we will consider only *S*-factor for the transition to the ground state of the $^8$Be nucleus i.e. the reaction: $^7$Li(p,γ$_0$)$^8$Be. One of the last experimental measurements of the *S*-factor of this reaction in the energy range from 100 keV to 1.5 MeV was made in the work [51].

Expressions given above in the first chapter are used for the *S*-factor calculations. Values: µ$_p$=2.792847 and µ($^7$Li)=3.256427 are accepted for magnetic momentum of proton and $^7$Li nucleus. The calculation results for the *S*-factor with the given above potentials at the energy range 5-800 keV (l.s.) are shown in Fig. 3. The *E*1 transition is shown by the dashed line, dotted line - *M*1 process, solid line - the sum of these processes. In the considered reaction the *M*1 transition like the *E*1 transition in the p$^3$H system [44] goes with change of the isospin Δ*T*=1, since the ground state of the $^8$Be nucleus has *T*=0 and resonance isospin in the $P_1$-wave of scattering equals 1.

The value of 0.50 keV b was obtained for the astrophysical *S*-factor at 5 keV (c.m.) for the transition to GS of the $^8$Be nucleus, where the *E*1 process gives the value of 0.48 keV b, which is in a good agreement with the data from [51]. The calculated and experimental *S*-factor values at the energy range 5-300 keV (l.s.) are given in Table 4. As it can be seen from Fig. 3 and Table 7, the value of the theoretical *S*-factor at the energy range 30-200 keV is almost constant and approximately equal to 0.41-0.43 keV b, which agrees with data of the work [51] for the energy range 100-200 keV practically within the experimental errors.

Let's compare some extrapolation results of different experimental data to zero energy. The value 0.25(5) keV b was obtained in work [52] and the value 0.40(3) keV b in work [53] on the basis of data [51]. Then in work [54] on the basis of new measurements of the total cross-sections of $^7$Li(p,γ$_0$)$^8$Be reaction at the energy range 40-100 keV the value 0.50(7) keV b was suggested which is in a good agreement with the obtained above value at the energy 5 keV.

**Table 4.** Calculated astrophysical *S*-factor of the reaction of p$^7$Li radiative capture at low energies and its comparison with the experimental data [51].

| $E_{lab}$, keV | $S_{exp}$, keV b [51] | $S_{E1}$, keV b | $S_{M1}$, keV b | $S_{E1+M1}$, keV b |
|---|---|---|---|---|
| 5.7 | --- | 0.48 | 0.02 | 0.50 |
| 29.7 | --- | 0.41 | 0.02 | 0.43 |
| 60.6 | --- | 0.39 | 0.02 | 0.41 |
| 98.3 | 0.41(3) | 0.39 | 0.03 | 0.42 |
| 198.3 | 0.40(2) | 0.37 | 0.06 | 0.43 |
| 298.6 | 0.49(2) | 0.36 | 0.16 | 0.52 |

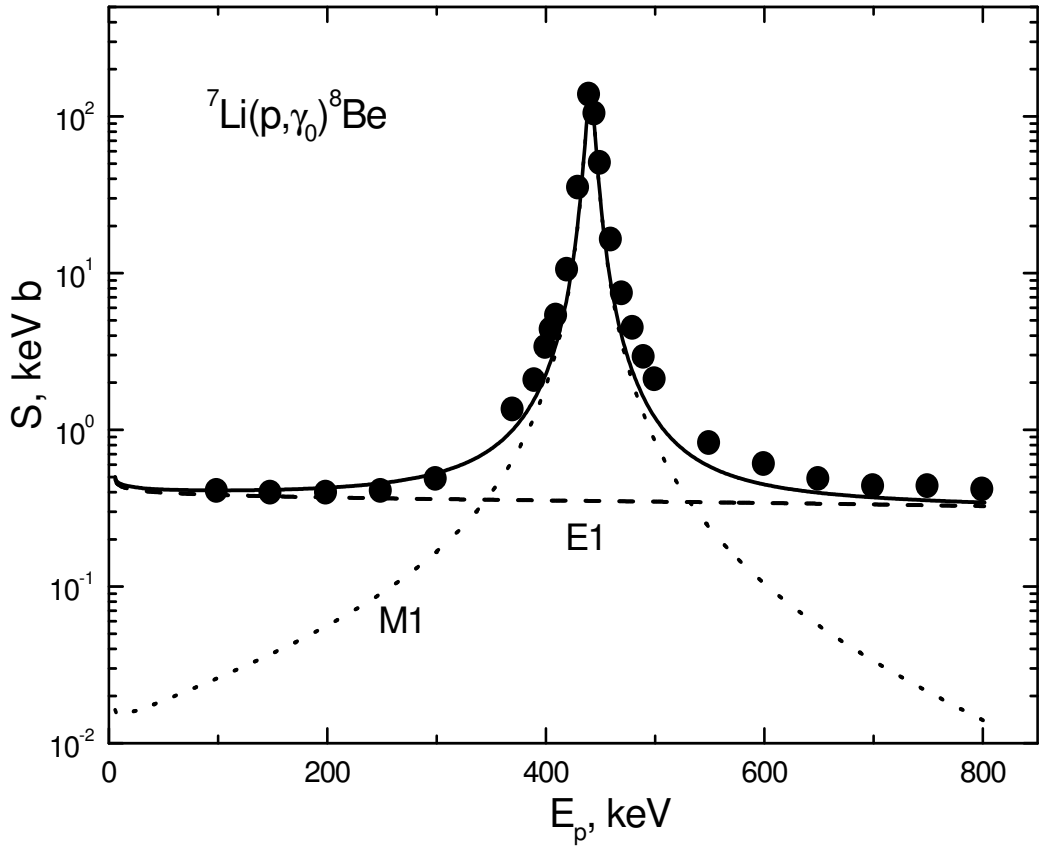

**Fig. 3.** Astrophysical *S*-factor of the reaction of p$^7$Li radiative capture. Dots: experimental data from work [51]. Lines: calculation results for different electromagnetic transitions with the potentials mentioned in the text.

It is interesting to look at the chronology of different works for determination of the astrophysical *S*-factor of the $^7$Li(p,$\gamma_0$)$^8$Be reaction. It was believed in 1992 that its value equals 0.25(5) keV b [52], the value 0.40(3) keV b [53] was obtained in 1997 on the basis of measurements made in 1995 [51] and the measurements in 1999 at lower energies led to the value of 0.50(7) keV b [54]. This chronology demonstrates well the constant increase in the value obtained for the astrophysical *S*-factor of $^7$Li(p,$\gamma_0$)$^8$Be reaction (two fold increase) as the energy of experimental measurements decreased.

Thus, $E$1 and $M$1 transitions from $S_1$ and $P_1$-wave of scattering to the ground bound state in the p$^7$Li channel of the $^8$Be nucleus were considered in the potential cluster model. It is possible to completely describe present day experimental data for the astrophysical $S$-factor at the energies up to 800 keV taking into account certain assumptions concerning the channel restructuring in the $^8$Be nucleus and to obtain its value for zero (5 keV) energy, which is in a good agreement with the latest experimental measurements.

## 5. Radiative p$^{12}$C capture

In this section we will consider the p$^{12}$C system and the process of proton radiative capture by the $^{12}$C nucleus at astrophysical energies. The new measurement of differential cross-sections of the elastic p$^{12}$C scattering at energies from 200 keV up to 1.1 MeV (c.m.) within the range of $10^0$-$170^0$ with 10% errors was carried out in works [55]. Further, the standard phase shift analysis was made and the potential of $S_{1/2}$-state of p$^{12}$C system was reconstructed in this paper on the basis of these measurements [39], and then the astrophysical $S$-factor at the energies down to 20 keV was considered in the frame of potential cluster model [56].

Let's start to immediate describing the results obtained, we would like to note that this process is the first thermonuclear reaction of the CNO-cycle which took place at a later stage of stellar evolution when a partial hydrogen burning occurred. As the hydrogen is burned the core of the star starts contracting which results in the increase in pressure and temperature in the star and along with the proton-proton chain the next chain triggers of thermonuclear processes, called CNO-cycle [2,4]. The p$^{12}$C radiative capture process is a part of the CNO thermonuclear cycle at low energies and gives a considerable contribution to energy efficiency of thermonuclear reactions [1,2].

### 5.1. Potentials of the p$^{12}$C interaction

As it was told before, on the basis of data [55], we have done in work [39] the phase shift analysis of the p$^{12}$C elastic scattering and the general view of $S_{1/2}$-phases are shown in Fig. 4, where black points - results of the phase shift analysis for the $S$-phase shift taking into account the $S$-wave only; open squares - results of the phase shift analysis for the S-phase shift taking into account $S$ and $P$-waves [39]; dashed line - result of work [57]; solid line - result calculated with potential (16). The scattering phase shifts obtained in such a way are used further for the construction of intercluster potentials and for calculations of the astrophysical $S$-factor. The existing experimental data of the astrophysical $S$-factor [9] indicates the presence of the narrow resonance with the width of about 32 keV at the energy 0.422 MeV (c.m.), which leads to the two-three order increase in the $S$-factor.

It is interesting to find out if there is a possibility to describe the resonance $S$-factor on the basis of the PCM with FS and with the classification of orbital states according to Young's schemes. The phase shift analysis of the new experimental data [55] of differential cross-sections of the elastic p$^{12}$C scattering at astrophysical energies [39], which we have shown above, allows constructing the potentials of the p$^{12}$C interaction for the phase shift analysis of the elastic scattering.

Let's examine the classification of orbital states according to Young's schemes in the p$^{12}$C system for the purposes of construction of the interaction potential. The possible orbital Young's schemes in the N=n$_1$+n$_2$ particle system can be defined in following way {1}×{444}={544} and {4441} [45]. The first of these schemes is consistent only with the orbital momentum $L$=0 and is forbidden, because $s$-shell cannot contain more than four nucleons [5]. The second scheme is allowed with orbital momenta 1 and 3 [45], the first of which corresponds to the ground bound state of the $^{13}$N nucleus with $J$=1/2$^-$. Therefore, in the potential of the $^2S_{1/2}$-wave there must be a forbidden bound state and $^2P$-wave should have only one allowed state at the energy of -1.9435 MeV [58].

For the calculations of photonuclear processes the nuclear part of the inter-cluster p$^{12}$C interaction is represented as (6) with the point-like Coulomb component. The potential of $^2S_{1/2}$-wave is constructed so as to describe correctly the corresponding partial phase shift of the elastic scattering, which has a well defined resonance at 0.457 MeV (l.s.).

Using the results of the phase shift analysis [39] the $^2S_{1/2}$-potential of the p$^{12}$C interaction with FS at energy $E_{FS}$=-25.5 MeV was obtained together with parameters:

$$V_S = -102.05 \text{ MeV}, \alpha_S = 0.195 \text{ fm}^{-2}, E_{FS} = -12.8 \text{ MeV}. \quad (16)$$

The results of calculation of $^2S_{1/2}$-phase shift with this potential are shown in Fig. 4 by the solid line.

The potential of the bound $^2P_{1/2}$-state has to reproduce correctly the bound energy of the $^{13}$N nucleus in the p$^{12}$C channel -1.9435 MeV [58] and reasonably describe the root-mean-square radius, which probably does not differ significantly from the radius of the $^{14}$N nucleus, which is equal to 2.560(11) fm [58]. As a result the following parameters were received:

$$V_{GS} = -144.492278 \text{ MeV}, \alpha_{GS} = 0.425 \text{ fm}^{-2}, \quad (17)$$

The potential gives the bound energy equals -1.943500 MeV and the root-mean-square radius $R_{ch}$=2.47 fm. We use the following values for the radii of proton and $^{12}$C: 0.8768(69) fm [12] and 2.472(15) fm [59]. The asymptotic constant $C_W$ with Whittaker asymptotics (8) was calculated for controlling behavior of WF of BS at long distances; its value in the range of 5-20 fm equals 1.36(1).

The variational method was used for an additional control of the bound energy calculations, which allowed to obtain the bound energy of -1.943499 MeV with the residual error being not more than 6 10$^{-14}$ for the first variant of BS potential (17) by using an independent variation of parameters and the grid having dimension 10. The asymptotic constant $C_W$ of the variational WF at distances of 5-17 fm remains at the level of 1.36(2). Its variational parameters are listed in [3]. The charge radius does not differ from the value obtained in FDM calculations.

For the real bound energy in this potential it is possible to use the value -1.9434995(5) MeV, i.e. the calculation error of finding bound energy is on the level of ±0.5 eV, because the variational energy decreases as the dimension of the basis increases and gives the upper limit of the true bound energy, but the finite-difference energy increases as the size of steps decreases and the number of steps increases.

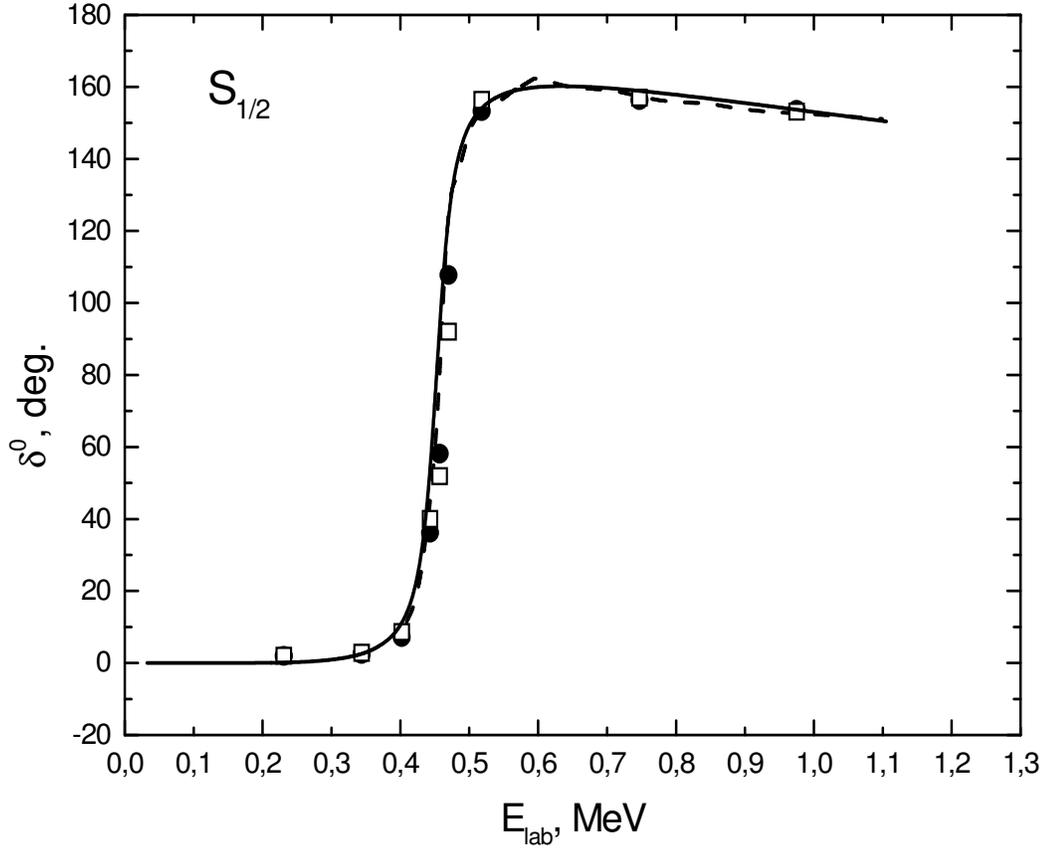

**Fig. 4.** $^2S$-phase shift of the elastic p$^{12}$C scattering at low energies. Black points - results of the phase shift analysis for the *S*-phase shift taking into account the S-wave only; open squares - results of the phase shift analysis for the *S*-phase shift taking into account *S* and *P*-waves [39]; dashed line - results of work [57]; solid lines - results calculated with potential (16).

*5.2. Astrophysical S-factor*

The *E*1(*L*) transition resulting from the orbital part of electric operator [10] is taken into account in present calculations of the process of radiative p$^{12}$C capture. The cross-sections of *E*2(*L*) and *MJ*(*L*) processes and the cross-sections depending on the spin part *EJ*(*S*), *M*2(*S*) turned out to be a few orders less. The electrical *E*1(*L*) transition in the p$^{12}$C→γ$^{13}$N process is possible between the doublet $^2S_{1/2}$ and $^2D_{3/2}$-states of scattering and the ground bound $^2P_{1/2}$-state of the $^{13}$N nucleus in the p$^{12}$C channel.

It should be noted that in all calculations the cross-section of the *E*1 electrical process due to transition from the doublet $^2D_{3/2}$-state of scattering to the ground bound $^2P_{1/2}$-state of the $^{13}$N nucleus is 4-5 orders less than the cross-section of the transition from $^2S_{1/2}$-state of scattering. Thus, the main contribution to the calculated *S*-factor of the p$^{12}$C→$^{13}$Nγ process is made by the *E*1 transition from the $^2S$-wave of scattering to the ground state of the $^{13}$N nucleus. The mass of proton was taken to be 1 in all calculations for the p$^{12}$C system.

The results of calculations of the *S*-factor of the radiative p$^{12}$C capture with the abovementioned potentials of $^2P_{1/2}$ and $^2S_{1/2}$-waves at energies from 20 keV to 1.0 MeV are shown together with experimental data from works [9,60] in Fig. 5 by the solid line. The value 1.52 keV b of the *S*-factor is received at energy 25 keV and the extrapolation of the *S*-factor experimental measurements to energy 25 keV gives: 1.45(20) keV b and $1.54^{+15}_{-10}$ keV b [58]. Though, in the range of 20-30 keV the *S*-factor value is practically constant and one can consider it as the *S*-factor value at zero energy with an error of about 0.02-0.03 keV b.

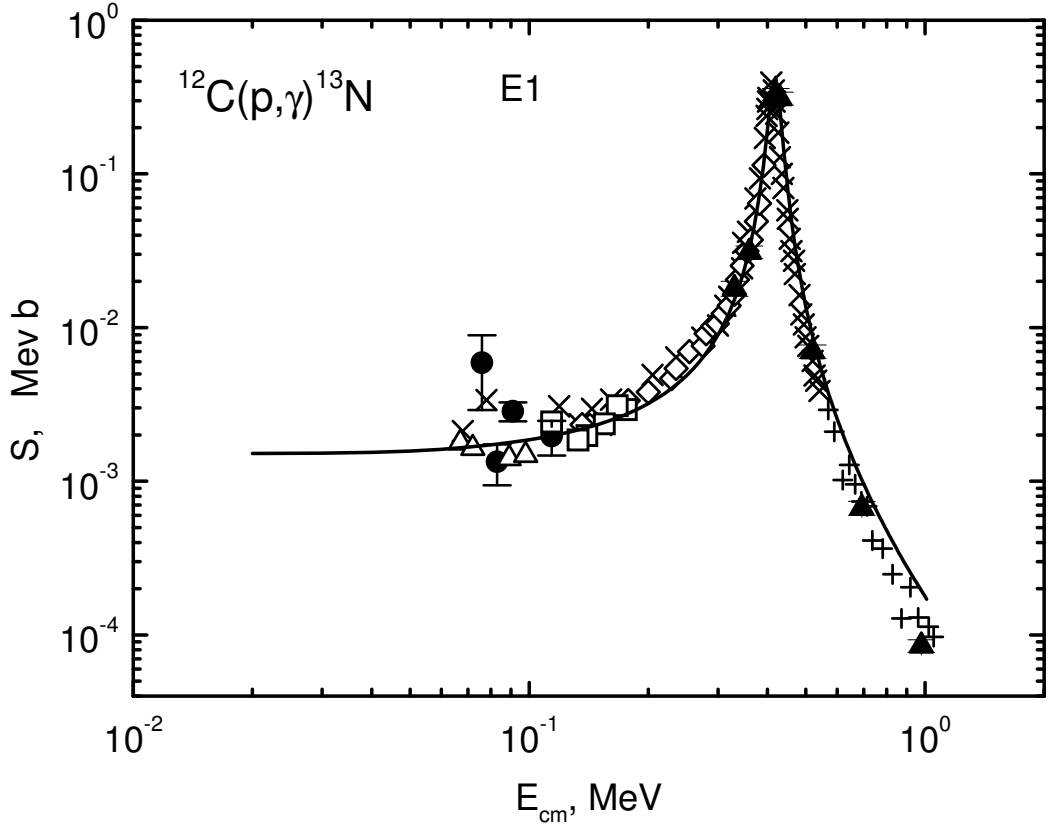

**Fig. 5.** Astrophysical *S*-factor of p$^{12}$C radiative capture at low energies. The experimental data specified as ×, •, □, +, ◊ and Δ are taken from review [9], triangles are from [60]. Line: calculations with potentials (16,17).

Thus, the given above potential with FS for the $^2S_{1/2}$-wave and the bound state without FS, which gives the correct bound energy, lead to the joint description of the resonance in the *S*-factor and the resonance in the $^2S_{1/2}$-phase of scattering.

At the same time, if we use the potentials of the $^2S_{1/2}$-wave with small depth and without forbidden states, for example, with parameters:

$V_S = -15.87$ MeV, $\alpha_S = 0.1$ fm$^{-2}$,
$V_S = -18.95$ MeV, $\alpha_S = 0.125$ fm$^{-2}$, (18)
$V_S = -21.91$ MeV, $\alpha_S = 0.15$ fm$^{-2}$,

then we can't obtain the correct description of the maximum of the *S*-factor of the radiative capture. It is impossible to describe the absolute value of the *S*-factor which for all variants of the scattering potentials (18) and the BS potentials is 2-3 times as much as the experimental maximum. At the same time, for all given depthless potentials of the form (18) the resonance behavior of the $^2S_{1/2}$-phase shift of scattering is well described. As the width of the $^2S_{1/2}$-potential decreases, i.e. the α value increases, the value of the *S*-factor maximum grows up, e.g. for the last variant of the $^2S_{1/2}$-scattering potential its value is approximately three times as much as the experimental value [56].

Thus, it is possible to describe the astrophysical *S*-factor and the $^2S_{1/2}$-phase shift of scattering in the resonance energy range 0.457 MeV (l.s.) on the basis of the PCM and the deep $^2S_{1/2}$-potential with the FS, and to receive the reasonable values for the charge radius and asymptotic constant. The depthless potentials of scattering do not lead to the joint description of the *S*-factor and the $^2S_{1/2}$-phase shift of scattering at any considered combinations of p$^{12}$C interactions [56].

## 6. Conclusion

The description of behavior of the *S*-factors in all considered systems at low energies may be viewed as certain evidence in favor of the potential approach in cluster model. The inter-cluster interactions including FS, which structure is determined by the orbital state classification according to the Young's schemes, are constructed on the basis of the phase shifts of the cluster elastic scattering, and each partial wave is described by its potential, for example of the Gaussian form, with certain parameters.

The splitting of the general interaction into the partial waves allows detailing its structure and the classification of the orbital states according to Young's schemes allows identifying the presence and the number of the forbidden states. It gives the possibility to find the number of nodes of the cluster relative motion WF of the GS nuclei and leads to a definite depth of the interaction allowing to avoid the discrete ambiguity of the potential depth as it is the case in the optical model.

The form of each partial phase shift of scattering can be correctly described only at the certain width of such potential and this is deliver us from the continuous potential ambiguity, which is also inherent to the conventional optical model. As a result, all the parameters of such a potential are fixed quite uniquely, and the "pure" according to Young's schemes interaction component allows describing the basic characteristics of the bound state of the lightest clusters generally correct, which is realized in the light atomic nuclei with a high probability. The requirement of true description of the BS characteristics, in such partial waves where they exist, is an additional criterion for the determination of intercluster potential parameters.

The formalism developed for obtaining intercluster potentials was applied here for description of nuclear photocapture reactions in the considered systems. The operator of electromagnetic transition for radiative capture processes, unlike other nuclear reactions due to strong interaction, is well known. Moreover, the interaction in the final state is absent in photocapture reactions and the interaction in the initial state is taken into account rather correctly based on the developed potential approach.

The calculation results of the *S*-factor of radiative p$^2$H capture at energies as low as 10 keV were carried out earlier in the framework of this cluster potential model, when only experimental data above 150÷200 keV were well-known and these results are in a good agreement with experimental the results in the range from 50 keV to 150÷200 keV that became available much later. Intrinsically, the behavior prediction of the *S*-factor of the p$^2$H radiative capture in this energy range was done.

However, all the above-mentioned relatively to methods of construction of intercluster potentials is correct provided that the phase shifts of scattering are obtained correctly from the experimental data of the elastic scattering. Up to present, for the majority of the lightest nuclear systems the phase shifts of scattering have been received with rather big errors, sometimes reaching 20-30%. This makes the construction of the exact potentials of the inter-cluster interaction very difficult and, finally, leads to significant ambiguities in the final results obtained in the potential cluster model.

In this connection the problem to raise the accuracy of experimental measurements of elastic scattering of light atomic nuclei at astrophysical energies and to perform a more accurate phase shift analysis is very urgent. The increase in the accuracy and using this semi-microscopic and semi-phenomenological model will allow making more definite conclusions regarding the mechanisms and conditions of thermonuclear reactions, as well as understanding better their nature in general.

**Acknowledgments**

Finally, the authors would like to express their deepest gratitude to Prof. Uzikov Yu.N. (JIRN, Dubna, Russia) and Prof. Blokhintsev L.D. (Moscow State University, Moscow, Russia) for the very important discussions of some parts of the work.

This work was supported in part by the V.G. Fessenkov's Astrophysical Institute NCSRT NSA RK with the help of Grant Program of Fundamental Research of the Ministry of Education and Science of the Republic of Kazakhstan.

**References**

[1]   W.A. Fowler, *Experimental and Theoretical Nuclear Astrophysics: the Quest for the Original of the Elements* (Nobel Lecture, Stockholm, 1983).
[2]   B.S. Ishkhanov, I.M. Kapitonov and I.A. Tutyn', *Nucleosynthesis in the Universe* (Moscow State University, Moscow, 1998) (in Russian); J.A. Peacock, *Cosmological Physics* (Cambridge University Press, Cambridge, 1999).
[3]   S.B. Dubovichenko, *Thermonuclear processes of the Universe* (APHI, Almaty, 2010). Available online at: http://nuclphys.sinp.msu.ru/thpu/index.html; http://arxiv.org/abs/1012.0877 (in Russian).
[4]   C.A. Barnes, D.D. Clayton and D.N. Schramm, *Essays in Nuclear Astrophysics Presented to William A. Fowler* (Cambridge University Press, Cambridge, 1982).
[5]   O.F. Nemets, V.G. Neudatchin, A.T. Rudchik, Y.F. Smirnov and Yu.M. Tchuvil'sky, *Nucleon Association in Atomic Nuclei and the Nuclear Reactions of the Many Nucleons Transfers* (Naukova dumka, Kiev, 1988) (in Russian).
[6]   V.I. Kukulin, V.G. Neudatchin and Yu.F. Smirnov, Fiz. Elem. Chastits At. Yadra **10**, 1236 (1979) [Sov. J. Part. Nucl. **10**, 1006 (1979)].
[7]   W.A. Fowler, G.R. Caughlan and B.A. Zimmerman, Annu. Rev. Astron. Astrophys. **13**, 69 (1975).
[8]   P.J. Mohr and B.N. Taylor, Rev. Mod. Phys. **77(1)**, 1 (2005).
[9]   C. Angulo, *et al.*, Nucl. Phys. A **656**, 3 (1999).
[10] S.B. Dubovichenko and A.V. Dzhazairov-Kakhramanov, Fiz. Elem. Chastits At. Yadra **28**, 1529 (1997) [Phys. Part. Nucl. **28**, 615 (1997)].
[11] D.A. Varshalovich, A.N. Moskalev and V.K. Khersonskii, *Quantum Theory of Angular Momentum* (World Scientific, Singapore, 1989).
[12] http://physics.nist.gov/cgi-bin/cuu/Value?mud|search_for=atomnuc!

[13] J.M. Eisenberg and W. Greiner, *Excitation mechanisms of the nucleus* (NorthHolland, Amsterdam, 1970).
[14] G.R. Plattner and R.D. Viollier, Nucl. Phys. A **365**, 8 (1981).
[15] G.I. Marchuk and V.E. Kolesov, *Application of Numerical Methods to Neutron Cross-Section Calculations* (Atomizdat, Moscow, 1970) (in Russian).
[16] V.G. Neudatchin, *et al.*, Phys. Rev. C **45**, 1512 (1992).
[17] P.E. Hodgson, *The Optical Model of Elastic Scattering* (Clarendon Press, Oxford, 1963).
[18] S.B. Dubovichenko, Yad. Fiz. **1**, 66 (2008) [Phys. Atom. Nucl. **71**, 65 (2008)].
[19] S.B. Dubovichenko, *Calculation methods for nuclear characteristics* (Complex, Almaty, 2006). Available online at: http://xxx.lanl.gov/abs/1006.4947 (in Russian).
[20] K.A. Snover, *Solar p-p chain and the $^7Be(p,\gamma)^8B$ S-factor* (CEPRA, NDM03, University of Washington, 1/6/2008).
[21] S.B. Dubovichenko and A.V. Dzhazairov-Kakhramanov, Euro. Phys. Jour. A **39**, 139 (2009); S.B. Dubovichenko, Izv. Vyssh. Uchebn. Zaved. Fiz. **2**, 28 (2011) [Russian Physics Journal **54**, 157 (2011)].
[22] S.B. Dubovichenko, Yad. Fiz. **58**, 1253 (1995) [Phys. Atom. Nucl. **58**, 1174 (1995)].
[23] P. Schmelzbach, *et al.*, Nucl. Phys. A **197**, 273 (1972); J. Arvieux, Nucl. Phys. A **102**, 513 (1967); J. Chauvin and J. Arvieux, Nucl. Phys. A **247**, 347 (1975); E. Huttel, *et al.*, Nucl. Phys. A **406**, 443 (1983).
[24] S.B. Dubovichenko and A.V. Dzhazairov-Kakhramanov, Yad. Fiz. **51**, 1541 (1990) [Sov. J. Nucl. Phys. **51**, 1971 (1990)].
[25] G.M. Griffiths, E.A. Larson and L.P. Robertson, Can. J. Phys. **40**, 402 (1962).
[26] L. Ma, *et al.*, Phys. Rev. C **55**, 588 (1997).
[27] G.J. Schimd, *et al.*, Phys. Rev. C **56**, 2565 (1997).
[28] C. Casella, H. Costantini, *et al*. (LUNA Collaboration), Nucl. Phys. A **706**, 203 (2002).
[29] D.R. Tilley, H.R. Weller and H.H. Hasan, Nucl. Phys. A **474**, 1 (1987).
[30] D.R. Tilley, H.R. Weller and G.M. Hale, Nucl. Phys. A **541**, 1 (1992).
[31] L.D. Blokhintsev, I. Borbely and E.I. Dolinskii, Fiz. Elem. Chastits At. Yadra **8**, 1189 (1977) [Sov. J. Part. Nucl. **8**, 485 (1977)].
[32] M. Bornard, *et al.*, Nucl. Phys. A **294**, 492 (1978).
[33] G.R. Platner, M. Bornard and R.D. Viollier, Phys. Rev. Lett. **39**, 127 (1977).
[34] A. Kievsky, *et al.*, Phys. Lett. B **406**, 292 (1997).
[35] Z. Ayer, *et al.*, Phys. Rev. C **52**, 2851 (1995).
[36] G.J. Schimd, *et al.*, Phys. Rev. Lett. **76**, 3088 (1996).
[37] G.J. Schimd, *et al.*, Phys. Rev. **52**, R1732 (1995).
[38] M. Viviani, R. Schiavilla and A. Kievsky, Phys. Rev. C **54**, 534 (1996).
[39] S.B. Dubovichenko, Izv. Vyssh. Uchebn. Zaved. Fiz. **11**, 21 (2008) [Russian Physics Journal **51**, 1136 (2008)].
[40] S.B. Dubovichenko, Yad. Fiz. **73**, 1573 (2010) [Phys. Atom. Nucl. **73**, 1526 (2010)].
[41] S.B. Dubovichenko, Izv. Vyssh. Uchebn. Zaved. Fiz. **12**, 30 (2010) [Russian Physics Journal **53**, 1254 (2010)].
[42] D.R. Tilley, *et al.*, Nucl. Phys. A **745**, 155 (2004).
[43] S.B. Dubovichenko, Izv. Vyssh. Uchebn. Zaved. Fiz. **3**, 68 (2009) [Russian Physics Journal **52**, 294 (2009)].

[44] S.B. Dubovichenko, V.G. Neudatchin, A.A. Sakharuk and Yu.F. Smirnov, Izv. Akad. Nauk SSSR, Ser. Fiz. **54**, 911 (1990); V.G. Neudatchin, A.A. Sakharuk and S.B. Dubovichenko, Few Body Systems **18**, 159 (1995).

[45] V.G. Neudatchin and Yu.F. Smirnov, *Nucleon associations in light nuclei* (Nauka, Moscow, 1969) (in Russian); V.I. Kukulin, V.G. Neudatchin, I.T. Obukhovsky and Yu.F. Smirnov, *Clusters as subsystems in light nuclei*, in: *Clustering Phenomena in Nuclei* edited by K. Wildermuth and P. Kramer (Vieweg, Braunschweig, 1983) vol. **3**, 1.

[46] C. Itzykson and M. Nauenberg, Rev. Mod. Phys. **38**, 95 (1966).

[47] S.B. Dubovichenko, *Characteristics of light atomic nuclei in the potential cluster model* (Daneker, Almaty, 2004). Available online at: http://xxx.lanl.gov/abs/1006.4944 (in Russian).

[48] W.D. Warters, W.A. Fowler and C.C. Lauritsen, Phys. Rev. **91**, 917 (1953).

[49] L. Brown, *et al.*, Nucl. Phys. A **206**, 353 (1973).

[50] D. R. Tilley, *et al.*, Nucl. Phys. A **708**, 3 (2002).

[51] D. Zahnow, *et al.*, Z. Phys. A **351**, 229 (1995).

[52] F. E. Cecil, *et al.*, Nucl. Phys. A **539**, 75 (1992).

[53] M. A. Godwin, *et al.*, Phys. Rev. C **56**, 1605 (1997).

[54] M. Spraker, *et al.*, Phys. Rev. C **61**, 015802 (1999).

[55] D.M. Zazulin, *et al.*, paper presented at the Sixth International Conference on Modern Problems of Nuclear Physics, Tashkent, 127 (2006); M.K. Baktybaev, *et al.*, paper presented at the Fourth Eurasian Conference on Nuclear Science and its Application, Baku, 56 (2006).

[56] S.B. Dubovichenko and A.V. Dzhazairov-Kakhramanov, Izv. Vyssh. Uchebn. Zaved. Fiz. **8**, 58 (2009) [Russian Physics Journal **52**, 833 (2009)].

[57] H.L. Jackson, *et al.*, Phys. Rev. **89**, 365 (1953); H.L. Jackson, *et al.*, Phys. Rev. **89**, 370 (1953).

[58] F. Ajzenberg-Selove, Nucl. Phys. A **523**, 1 (1991).

[59] F. Ajzenberg-Selove, Nucl. Phys. A **506**, 1 (1990).

[60] N. Burtebaev, S.B. Igamov, R.J. Peterson, R. Yarmukhamedov and D.M. Zazulin, Phys. Rev. C **78**, 035802-1 (2008).